\documentclass[prl,floatfix,twocolumn]{revtex4}
\usepackage{graphicx}
\begin{document}
\title{Negative Magnetoresistance in Andreev Interferometers}
\author{W. Belzig}
\affiliation{Department of Physics and Astronomy, University of Basel,
Klingelbergstr. 82, 4056 Basel, Switzerland}
\author{R. Shaikhaidarov}
\author{V.~V. Petrashov}
\affiliation{Department of Physics, Royal Holloway, University of London,
  Egham, Surrey TW20 OEX, United Kingdom} 
\author{Yu.~V. Nazarov}
\affiliation{Department of Applied Physics, Delft University of
  Technology, Lorentzweg 1, 2628CJ Delft, Netherlands}
\begin{abstract}
  We consider transport in a diffusive cross-shaped Andreev
  interferometer geometry, both theoretically and experimentally. A
  strong and unexpected modulation of the conductance with the
  superconducting phase is found.  In particular, a reversed resistance
  vs.  phase difference is predicted, where the resistance is {\em
    decreased} by a phase gradient. A comparison of our quasiclassical
  calculation with experimental data shows quantitative agreement if we
  account for diamagnetic screening.
\end{abstract}
\pacs{74.50.+r, 73.23.-b, 75.75.+a}

\maketitle

Quantum mechanical interference effects can be ideally studied in
heterostructures of normal metals and superconductors. Large conductance
oscillations in structures containing loops threaded by a magnetic flux
have been predicted and observed
\cite{petrashov1,pothier,petrashov2,vanwees,courtois,volkov,stoof:96,GWZ}.
The origin of these oscillation is the phase coherence between Andreev
coupled electron-hole pairs, which can be maintained over large
distances of the order of $\mu$m, making their observation in
microstructured devices accessible
\cite{VZK,raimondi,suplattrev,pannetier}. Since the origin of these
oscillation are superconducting correlations, they are usually
suppressed by a magnetic field, i.e.  these devices show a positive
magnetoresistance. This trend can be reversed by means of tunnel
junctions, in which case the conductance is strongly reduced due to a
suppression of the low energy density of states leads.  Obviously, the
suppression of superconducting correlations then leads to a decreased
resistance \cite{antonov}. Thus, one would expect that a magnetic field
drives the resistance towards its normal state value, no matter in what
direction this actually is.

Here we report on a theoretical calculation and an experimental
measurement of the magnetotransport of a purely diffusive
heterostructure, which shows a negative magnetoresistance due to an
subtle interference effect.  This is surprising, since generally one
would expect, that both finite energies and a phase gradient would lead
to a destructive interference, diminishing the proximity effect.  Thus,
although the conductance shows the so-called reentrance effect as
function of temperature, one would not expect a phase gradient to
reverse the temperature effect. In the present Letter we study a
geometry, in which a separation of the energy scales responsible for
these two effects result in a negative magnetoresistance.

We will first describe the theoretical approach, which predicts a
negative magnetoresistance for small magnetic flux.  To obtain the
experimentally observed fully inverted resistance oscillations, we will
have to include diamagnetic screening in the superconducting loop
imposing the phase difference. This allows us to obtain excellent
agreement using parameters determined from the experiment independently.
We note that this agreement is contrary to the claim in
Ref.~\cite{shaikaidarov1}, in which a similar experimental layout was
studied. It was however claimed, that the experimental observations can
not be explained within the 'standard theory' of the mesoscopic
diffusive proximity effect. Here, we show that these observations are in
perfect agreement with the 'standard' quasiclassical theory.

\begin{figure}
  \begin{center}
    \includegraphics[width=\columnwidth]{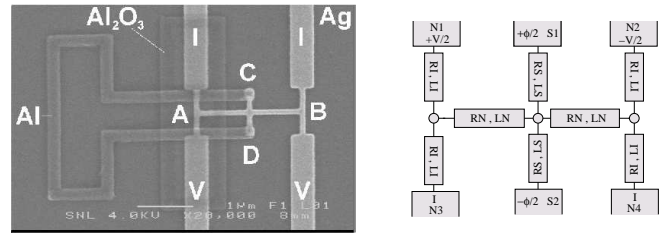}
    \caption[]{Geometry of experimental setup and the theoretical
      model. The left picture shows a micrograph of the experimental
      layout. The resistance of the horizontal Ag wire is measured in
      4-terminal configuration by current(I)- and voltage(V)-probes
      attached to points A and B. In the middle two superconducting
      mirrors C and D made from Al are attached by Ag wires. The
      superconducting mirrors are connected to loop threaded
      by a magnetic flux, which allows to vary the phase difference
      between C and D. Theoretically we model the experimental setup by
      the structure shown in the right panel. The resistors $R_S$, $R_N$
      and $R_I$ model the respective Ag-wires of length $L_S$, $L_N$ and
      $L_I$.}
    \label{fig:system1}
  \end{center}
\end{figure}

The system we study is shown in Fig.~\ref{fig:system1}. The left part
shows the experimental setup and the right shows the theoretical model
structure. The resistance of the vertical piece of normal metal is to be
measured.  Each part (left and right) of this vertical wire has a
resistance $R_N$ and a length $L_N$. The reservoirs inducing
superconductivity (S1 and S2) are attached to the middle of the wire by
normal resistors of length $L_S$ and resistance $R_S$. The actual
measurement of the resistance is performed in a
4-terminal geometry by attaching current and voltage probes (N1-N4)
through resistors of length $L_I$ and resistance $R_I$. Note, that these
probes have to be included into the modeling in contrast to normal
systems. In a proximity effect structure, the properties of these probes
matter, since they modify the equilibrium properties (i.e. the density
of states and therefore the energy- and space-dependent conductivity of
the system).

The calculation of the resistance of the structure depicted in
Fig.\ref{fig:system1} has been performed numerically, along the lines of
Ref.~\cite{nazarov:99,stoof:96} using the quasiclassical
formalism\cite{quasicl}. From the spectral Usadel equation the local
energy-dependent conductivity is extracted via
\begin{equation}
  \label{eq:local_cond}
  \sigma(E,x)=\frac{\sigma_N}{4}\text{Tr}\left[\hat 1 -\hat\tau_z\hat
    g_R\hat\tau_z\hat g_A\right]\;.
\end{equation}
Here $\sigma_N$ is the normal state conductivity and $\hat
g_{R,A}$ are retarded and advanced Nambu Green's functions
obeying the Usadel equation (see Ref.~\cite{nazarov:99} for a
definition).  The solution of the kinetic equation yields for the
measured 4-terminal resistance
\begin{equation}
  \label{eq:resistance}
  R^{-1}=\int \frac{dE}{4T\cosh^{2}\left(\frac{E}{2T}\right)} 
  \left(\int_0^{L_N}
  \frac{dx}{\sigma(E,x)} \right)^{-1}\,.
\end{equation}
Similarly we can find the supercurrent in equilibrium as
\begin{equation}
  \label{eq:super}
  I_{S}=\frac{\sigma}{8}\int dE \textrm{Im}\textrm{Tr}
  \left[\hat\tau_z \hat g_R \frac{\partial}{\partial y} \hat g_R
  \right]
  \tanh\left(\frac{E}{2T}\right)\,.
\end{equation}
These expressions show, that the 4-Terminal configuration indeed
measures the resistance of the vertical branch only. However, due to the
proximity effect the spectral conductivity $\sigma(E,x)$ depends also on
the properties of the current and voltage leads. As a consequence this
will also lead to a strongly increased supercurrent for temperatures at
which these branches are probed, as we demonstrate below.

\begin{figure}
  \begin{center}
    \includegraphics[width=\columnwidth,clip=true]{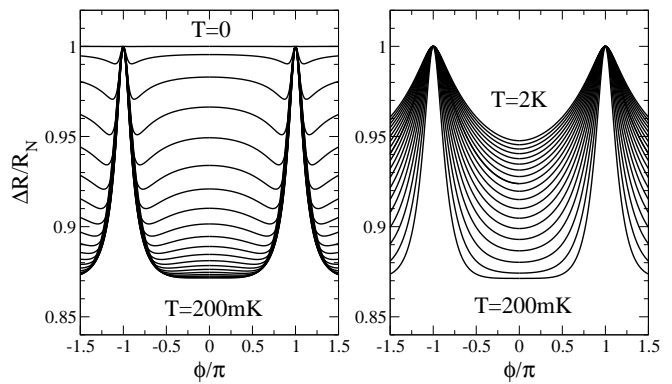}
    \caption{Resistance versus phase difference. 
      Right plot: resistance oscillations for temperatures between 200mK
      and 2K in steps of 100mK.  Left plot: resistance oscillations for
      temperatures from 0 to 200mK in steps of 10mK.  The form of the
      oscillations at temperatures above 200mK are as expected, i.e. the
      resistance monotonously increases between 0 to $\pi$. Below 200mK
      the phase dependence changes qualitatively. In particular for
      small phase difference the magnetoresistance is negative.}
    \label{fig:resist1}
  \end{center}
\end{figure}

We now discuss the concrete theoretical results. The parameters,
that we have chosen correspond to the experimental setup shown in the
left panel of Fig.~\ref{fig:system1}. The dimensions are $L_N=1000$nm,
$L_S= 225$nm and $L_I=800$nm, where $L_I$ is taken a little bit
larger than in the experiment to account for the proximity effect into
the wider ("lead") regions of the experiment.  For the connectors to the
superconductor we have taken the distance between the central point and
the closest point of the superconductor.  Taking the experimental value
of the diffusion constant $D\approx 80$cm$^2/$s, we find for the
Thouless energy $ E_{N}=\hbar D/L_N^2= 57$mK$k_B$. The branch connecting
the two superconductors directly has correspondingly $E_{S}=\hbar D/4
L_S^2= 280$mK$k_B$ (we took the full length, since this is mainly used
in literature to characterize SNS junctions). In the calculation we have
assumed a temperature independent superconducting gap corresponding to
$\Delta/k_B=2$K.

Let us now turn to the phase dependent resistance of the vertical
branch, measured in 4-point configuration through the leads N1-N4. The
resistance oscillations for a large range of temperatures are shown in
Fig.~\ref{fig:resist1}. Following the temperature dependence of the
resistance for zero phase difference, we observe the usual reentrance
behaviour, with a maximal suppression of about 13\% at a temperature
around 200mK. At a phase difference of $\pi$ (or any odd multiple), the
proximity effect is completely suppressed for all temperatures. At
temperatures above 200mK the resistance vs. phase oscillation are
similar to that predicted previously \cite{volkov,stoof:96,GWZ} and
found experimentally\cite{courtois,petrashov1}. However, below 200mK the
oscillations change qualitatively. Whereas the resistance increases
monotonically starting from zero phase difference for the larger
temperatures, it {\em decreases} first for the lower temperatures. Only
close to half-integer phases the resistance increases again. At
half-integer flux the resistance is always equal to the normal state
value, as it should be.

\begin{figure}[t]
  \begin{center}
    \includegraphics[width=\columnwidth]{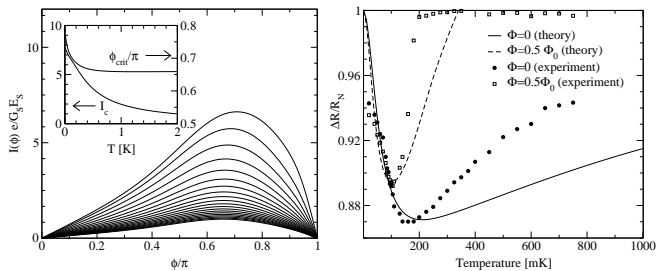}
    \caption{Left panel: Supercurrent phase relation (main) and critical
      current/critical phase (inset). The non-sinusoidal current phase
      relation is plotted for temperatures from 100mK to 1K in steps of
      100mK (from bottom to top). For phase differences around 3$\pi$/4
      the increase is strongest. The inset shows the temperature
      dependence of the critical current, together with the phase at
      which it is reached.  Both quantities show a strong variation
      below ~1K. Right panel: resistance versus temperature for integer
      and half-integer external flux. The lines are theoretical results
      and the symbols correspond to the experimental data. For integer flux
      both, experiment and theory, show the usual reentrance behaviour.
      For half-integer flux, the resistance decreases from the normal
      state resistance below a temperature of about 400mK. At lower
      temperatures it drops even below the resistance for integer flux.
      We attribute the saturation of the experimental resistance at very
      low temperatures to heating effects. The best fit of the
      resistance for half-integer flux was obtained for a screening
      parameter $\gamma=0.005 e/G_NE_N$.}
    \label{fig:zero_pi_resistance}
  \end{center}
\end{figure}

Obviously, such a strong change of the phase-dependence should also
affect other transport properties, for example the temperature- and the
phase-dependence of the supercurrent.  The theoretical results are shown
in the inset of the left panel of Fig.~\ref{fig:zero_pi_resistance}.
Note that, for simplicity, we have not included the temperature
dependence of the superconducting gap.  Therefore, the supercurrent is
finite above the experimental critical temperature of $\approx 1.4K$.
The critical current strongly increases below a temperature close to
$E_S$. At higher temperatures the critical current depends only weakly
on the temperature (the exponential decrease $\sim \exp(-\sqrt{k_B
  T/E_S})$ sets in only at much larger temperatures). The strong
increase is related to the influence of the vertical branch. As
superconducting correlation become strong in this branch, the pair
breaking influence of normal lead gets effectively suppressed.
Accordingly the supercurrent increases drastically. The
phase-difference, for which the critical current is reached, shifts from
$0.65\pi$ up to $0.8\pi$ in the same temperature interval.  In the left
panel of Fig.~\ref{fig:zero_pi_resistance} we show the current-phase
relation in an temperature interval between 100mK and 2K. It is clearly
non-sinusoidal and also changes qualitatively at low temperatures.

\begin{figure}[t]
  \begin{center}
    \includegraphics[width=\columnwidth,clip=true]{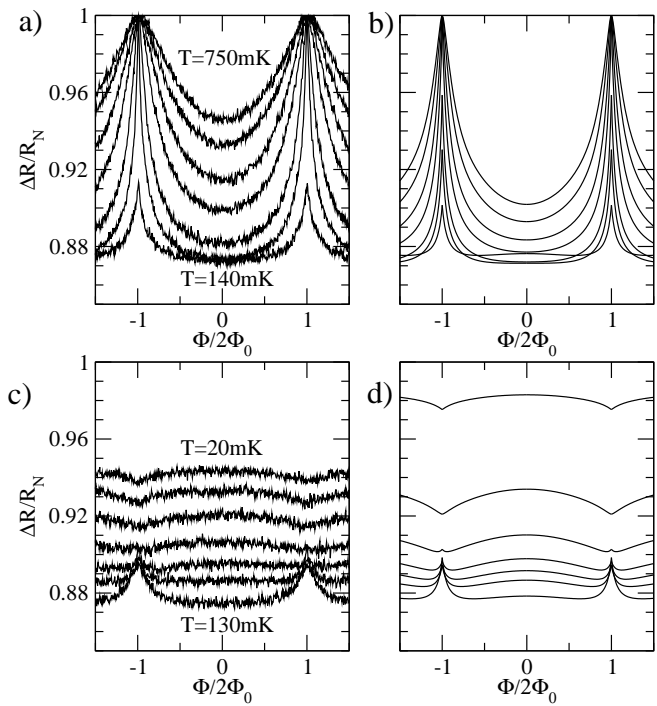}
    \caption[]{Resistance oscillations including screening. We compare the
      theoretical predictions (right column) with the experimental
      results (left column) using the screening parameter, determined
      from the fit in Fig.~\ref{fig:zero_pi_resistance}. In the upper
      row the resistance oscillations are compared for temperatures of
      750, 600, 450, 350, 250, 200, 140 mK; in the lower row for 130,
      105, 95, 85, 70, 50, 20 mK.  The agreement is very satisfactory,
      although not all details coincide.}
    \label{fig:resist_phase}
  \end{center}
\end{figure}

Summarizing these theoretical results, we have observed a rather
surprising phase-dependence of transport properties in a multi-terminal
diffusive heterostructure. However, the experimental results, shown in
the left column of Fig.~\ref{fig:resist_phase}, are quite different. The
reason, as we will quantify below, is the diamagnetic screening in the
superconducting loop. The applied magnetic flux induces a supercurrent
in the loop, which screens the magnetic flux.  Therefore, the phase
difference between the superconducting reservoirs is not directly given
by the applied magnetic flux. This effect depends on the self-inductance
of loop. We will show below, that accounting of this leads to a fully
inverted phase dependence and a very good agreement of experimental
results and theoretical calculations.

To address the screening effect we consider a superconducting loop interrupted
by a weak link. For the system depicted in Fig.~\ref{fig:system1} the weak
link is the normal metal structure. We therefore have to assume that the
entire phase-drop occurs over the normal metal, which is equivalent to say that
the critical current in the normal metal (induced by proximity) is smaller
than the critical current in the superconducting ring connecting the two
reservoirs S1 and S2. The energy of a superconducting loop, containing a weak
link reads,
\begin{equation}
  \label{eq:loop_energy}
  E(\varphi)=\frac{1}{2} L I(\varphi)^2 + E_{\text{J}} (\varphi)\,.
\end{equation}
Here, $L$ is the selfinductance of the loop, which depends on the
geometry, and $E_{\text{J}}(\varphi)$ is the Josephson energy of the
weak link.  It has to be determined from the integration of the
current-phase relation $I(\varphi)$.  Fluxoid quantization requires that
$\varphi=2\pi \Phi_{\text{tot}}/\Phi_0$ (we restrict ourselves here to
$\varphi\in[0,\pi]$, for simplicity).  The total flux
$\Phi_{\text{tot}}$ is the sum of the externally applied flux
$\Phi_{\text{x}}$ and the flux created by the supercurrent
$\Phi_{\text{ind}}=LI(\varphi)$. Introducing
$\varphi_x=2\pi\Phi_x/\Phi_0$ and minimizing (\ref{eq:loop_energy}) with
respect to $\varphi$ we obtain an equation, which determines the phase
difference as function of the external flux
\begin{equation}
  \label{eq:mini-cond}
  \varphi_x-\varphi = \frac{2e}{\hbar} L I(\varphi) \equiv
  \gamma I(\varphi)\,.
\end{equation}
In the limit of vanishing selfinductance ($L\to 0$) the phase difference
across the junction is equal to the externally applied flux. This is
usually the desired result. On the other hand, if the left hand side of
Eq.~(\ref{eq:mini-cond}) is not negligible, the phase at the junction
differs from the applied flux. If e.g. the current is
$I(\varphi)=I_c\sin\varphi$ and $(2e/\hbar) L I_c\gg 1$, we find
$\varphi\approx \varphi_x/(2eLI_c/\hbar) \ll \varphi_x$. The phase is
always much smaller than the applied flux. This holds as long as the
applied flux is less than $\pi$. If the external flux exceeds this
value, other solutions become important and the phase jumps. The main
consequence of these jumps is, that certain intervals of phases around
odd multiples of $\pi$ can not be reached by modulation of the external
flux.  Below, we will fully account for the screening, by calculating
the full current phase-relation for each temperature. Then the solution
of Eq.~(\ref{eq:mini-cond}) determines the actual phase difference.

As discussed previously accounting for the screening effect renders a
certain interval of phase differences unobservable. Thus, we expect that
screening suppresses the resistance at half integer external flux, since
the actual phase difference differs from an odd multiple of $\pi$.  The
screening effect is directly seen in the temperature dependence of the
resistance for half-integer flux. This is depicted in the right panel
Fig.~\ref{fig:zero_pi_resistance} together with the zero-flux
resistance.  The value of the selfinductance parameter $\gamma$ is
chosen to match the experimentally achieved minimal resistance around
100mK. Note, that without screening the resistance at half-integer flux
would be always equal to the normal state resistance. Taking screening
into account strongly suppresses the resistance below 400mK with a
maximal suppression of $\sim$7\%. At lower temperatures, the
resistance reenters, similar as the zero-flux resistance, back to the
normal states resistance at zero temperature. As can be seen the
resistance for half-integer flux is also slightly below the zero-flux
resistance, indicating a reversal of the resistance vs.  flux
oscillations. This is in agreement with the experimental data. The
deviations at high temperatures can be attributed to the neglection of
the temperature dependence of the superconducting gap.

To confirm, that the theoretical results are fully consistent with the
experimental data, we compare the resistance oscillations for various
temperatures in Fig.~\ref{fig:resist_phase} using the screening
parameter, determined from the fit in Fig.~\ref{fig:zero_pi_resistance}.
The similarity to the experimental curves is quite striking, although
not all details coincide. However, it is clearly seen, that the
screening effect can lead to the inverted resistance oscillations
("$\pi$-shift") at low temperatures, i.~e. the system has a negative
magnetoresistance.

Let us finally comment on the numerical value of the screening
parameter, $\gamma=0.005 e/G_N E_N$. Using the experimental values
$R_N\approx 3\Omega$ and $E_N=57$mK$k_B$, we calculate for the
selfinductance of the loop $L\approx 2$pH.  On the other hand, we find for
a circular loop of radius $R=2\mu$m and cross section
$W^2=20\times 100 $nm$^2$ a  self-inductance $L_{\text{loop}}=\mu_0
R(\ln(8R/W)-7/4)=9$pH. Although the selfinductance of a loop with the
experimental geometry is not easy to determine, this rough estimate
agrees very well with the selfinductance determined from our fit.

In conclusion we have shown, that a diffusive Andreev interferometer can
show a negative magnetoresistance and a non-sinusoidal
supercurrent-phase relation. This effect has been calculated
theoretically and found experimentally.  We obtained excellent agreement
with experimental data, if the diamagnetic screening in the
superconducting loop imposing the phase difference is taken into
account.  This eventually leads to a fully inverted ('$\pi$-shifted')
resistance oscillations.

\acknowledgements W.~B. was supported by the Swiss NSF and the NCCR
Nanoscience.


\begin{thebibliography}{99}

\bibitem{petrashov1} V.~T. Petrashov, V.~N. Antonov, P. Delsing, and
  T. Claeson, JETP Lett. {\bf 60}, 606 (1994).

\bibitem{pothier} H. Pothier, S. Gueron, D. Esteve, and M.~H. Devoret,
  Phys. Rev. Lett {\bf 73}, 2488 (1994).

\bibitem{petrashov2} V.~T. Petrashov, V.~N. Antonov, P. Delsing, and
  T. Claeson, Phys. Rev. Lett. {\bf 74}, 5268 (1995).
  
\bibitem{vanwees} A. Dimoulas, J.~P. Heida, B.~J.~v. Wees, T.~M. Klapwijk,
  W.~v.~d. Graaf, and G. Borghs, Phys. Rev. Lett. {\bf 74}, 602 (1995).
  
\bibitem{courtois} P. Charlat, H. Courtois, , Ph. Gandit, D. Mailly, A.
  F. Volkov, and B. Pannetier, Phys. Rev. Lett. {\bf 77}, 4950 (1996).
  
\bibitem{volkov} A. Volkov, N. Allsopp, C.~J. Lambert, J. Phys.:
  Condens. Matter {\bf 8}, L45 (1996).

\bibitem{stoof:96} Yu.~V. Nazarov and T.~H. Stoof, Phys. Rev. Lett. {\bf
    76}, 823 (1996); T.~H. Stoof and Yu.~V. Nazarov, Phys. Rev. B {\bf
    54}, R772 (1996).

\bibitem{GWZ} A.~A. Golubov, F.~K. Wilhelm, and A.~D. Zaikin, Phys. Rev. B
  {\bf 55}, 1123 (1997).

\bibitem{VZK} A.~F. Volkov, A.~V. Zaitsev, and T.~M. Klapwijk, Physica C
  {\bf 59}, 21 (1993).

\bibitem{raimondi} C.~J. Lambert and R. Raimondi,
  J. Phys. Cond. Mat. {\bf 10}, 901 (1998).

\bibitem{suplattrev} W. Belzig, F.~K. Wilhelm, C. Bruder,G.\
  Sch\"on, and A.~D. Zaikin, Superlattices Microst. {\bf 25}, 1251
  (1999).

\bibitem{pannetier} B. Pannetier and H. Courtois , 
  J. of Low Temp. Phys. {\bf 118}, 599 (2000)
  
\bibitem{antonov} V.~N. Antonov, H. Takayanagi, F.~K. Wilhelm, and A.~D.
  Zaikin Europhys. Lett. {\bf 50}, 250 (2000).

\bibitem{shaikaidarov1} A. Kadigrobov, L.~Y. Gorelik , R.~I. Shekhter, M.
  Jonson, R.~Sh. Shaikhaidarov, V.~T. Petrashov, P. Delsing and T. Claeson,
  Phys. Rev. B {\bf 60}, 14589 (1999).
  
\bibitem{nazarov:99} Yu.~V.  Nazarov, Superlattices Microst. {\bf 25},
  1221 (1999).

\bibitem{quasicl}
  G. Eilenberger, Z. Phys. {\bf 214}, 195 (1968);
  A.~I. Larkin and Yu.~N. Ovchinnikov,
  Sov. Phys. JETP {\bf 26}, 1200 (1968);  
  K.~D. Usadel,  Phys. Rev. Lett. {\bf 25}, 507 (1970).

\end{thebibliography}
\end{document}